\documentstyle[12pt,epsf]{article}
\def\beq{\begin{eqnarray}}   \def\eeq{\end{eqnarray}}

\begin{document}
\thispagestyle{empty}
\begin{flushright}
NYU-TH/00/03/05\\
\end{flushright}

\vspace{0.1in}
\begin{center}
\bigskip\bigskip
{\Large \bf A Comment  on Brane Bending and Ghosts in 
\\  \vspace{0.1cm} 
Theories
with Infinite Extra Dimensions}

\vspace{0.3in}

{ Gia Dvali, Gregory Gabadadze, and  Massimo  Porrati}
\vspace{0.1in}

{\baselineskip=14pt \it 
Department of Physics, New York University, New York, NY 10003 } \\
\vspace{0.2in}
\end{center}

\vspace{0.9cm}
\begin{center}
{\bf Abstract}
\end{center} 
\vspace{0.1in}

Theories with infinite volume extra dimensions open exciting opportunities
for particle physics. We argued recently that along with 
attractive features there are phenomenological difficulties in  this 
class of  models.
In fact, there is no graviton zero-mode in this case and 4D gravity is
obtained by means of  continuum bulk modes. 
These modes have additional degrees
of freedom which do not decouple at low energies and lead to 
inconsistent predictions for light  bending and the 
precession of Mercury's perihelion.
In a recent papers, [hep-th/0003020] and [hep-th/0003045]
the authors made  use of brane bending in order
to cancel the unwanted physical polarization of gravitons.
In this note we point out  that this mechanism does not
solve the problem since it uses a {\it ghost} 
which  cancels the extra degrees of freedom.   
In order to have a consistent model  the ghost should be eliminated. 
As soon as this is done, 4D gravity becomes unconventional and 
contradicts General Relativity.
New mechanisms are needed to cure these models. 
We also comment on the possible
decoupling of the ghost at large distances
due to an apparent flat-5D nature of space-time
and on the link between the presence of ghosts and the violation
of positive-energy conditions.

\newpage

Theories with infinite volume extra dimensions 
open exciting opportunities
for particle physics. The following 
5D warped metric may serve as a good example of this class of models:
\beq
ds^2~=~ A(y) \eta_{\mu\nu}~ dx^\mu dx^\nu~-~
dy^2~,
\eeq 
where the warp factor $A(y)$ tends to a {\it nonzero} 
constant at $\pm \infty$. 
A brane setup which realizes this  was recently proposed in 
Ref. \cite {Rubakov1}. It was argued in Refs. \cite {DGP} and \cite
{Witten} that these models are very attractive since they 
could give new insights into bulk supersymmetry and the 
cosmological constant problem.
Regretfully, as they stand right now,  
these theories face two serious challenges:   
to reproduce the  correct four-dimensional Einstein limit 
without invoking  ghost states \cite{DGP}, and to satisfy  a 
weak energy positivity condition \cite {Witten}.

The aim of the present note is to respond to the criticism of
\cite{Csaki2}  and \cite {Rubakov2} regarding the first issue.

Let us first recall the arguments of Ref. \cite{DGP}. 
The work was based on the following assumptions:

I) The theory is self-consistent, in the sense that it has no {\it
unconventional} or unphysical states,  such as ghosts;

II) 5D gravity couples universally to the energy-momentum tensor
   $T_{\mu\nu}$.

We argue that in \cite {Csaki2} and \cite {Rubakov2}
the condition (I) is relaxed. 

In models with infinite extra dimensions, 
differently form the Randall-Sundrum (RS) model
\cite {RS}, there is no localized 4D spin-2 or spin-0 zero-mode.    
The only relevant physical degrees of freedom are 4D massive spin-2
gravitons. As a result, 4D  gravity   
is obtained by exchanging  a metastable graviton
\cite {Rubakov1,Csaki1,DGP}. 
This is equivalent to the exchange of a continuum of massive spin-2 bulk
states.
Each of the continuum  states, from the 4D point of view, has 
strictly 5 physical degrees of freedom.
They can be
conveniently
decomposed  as:
2 from the 4D massless graviton, 2 from  the ``graviphoton'' and 1 from a 
``graviscalar''.
Two of these, coming from the ``graviphotons'' are not relevant for 
matter localized on the brane. 
Graviscalars contribute to physical processes \cite {Veltman,Zakharov}. 
These extra scalar degrees of freedom lead to deviations from
the standard predictions of Einstein's theory 
\cite {Veltman,Zakharov,DGP} since the tensor structures of massive
and massless graviton propagators are different:
\beq
\left( {1\over
2}(\eta^{\mu\alpha}\eta^{\nu\beta} + \eta^{\mu\beta}\eta^{\nu\alpha}) - 
{1\over
3}\eta^{\mu\nu}\eta^{\alpha\beta}+{\cal O}(p) \right) && {\rm massive}~;
\nonumber \\
\left( {1\over
2}(\eta^{\mu\alpha}\eta^{\nu\beta} + \eta^{\mu\beta}\eta^{\nu\alpha}) - 
{1\over
2}\eta^{\mu\nu}\eta^{\alpha\beta}+{\cal O}(p) \right)  && {\rm massless}~.
\eeq
Under assumptions (I) and (II), the 4D gravitational interactions
are completely determined by the exchange of bulk gravitons. As we
emphasized above, 
from the 4D point of view,  
these are just massive spin-2 states,  with 5 degrees of freedom for each
of them.  The effective 4D gravity in \cite {Rubakov1} is obtained by 
summing up these states. Therefore,  it is clear that the degrees of 
freedom do not match with those of 4D General Relativity and lead to 
unacceptable predictions \cite {DGP}. In other words, there is an additional
scalar degree of freedom in the 4D world obtained in \cite {Rubakov1}.  
Note that our arguments are
very general and are based only on the assumption of unbroken 
4D general covariance.

The way to evade  this result is to compensate 
the extra scalar with a  ghost state.  
Clearly, if one introduces 
unconventional states, such as ghosts 
\cite {Csaki2,Rubakov2},   
the results of \cite {DGP} are modified, but
then it is hard to make sense of the theory 
(see discussions below).

It was suggested in Ref. \cite {Csaki2} that the
unwanted polarizations
are canceled if brane bending is taken into account. 
The question is how one can 
reconcile this claim with the 4D arguments presented above?
The only way is by relaxing the assumption (I) of Ref. \cite 
{DGP} and allowing for a ghost state in the theory. 

To see that
this is indeed the case in \cite {Csaki2} let us recall that 
the brane bending
studied in \cite {gt} and \cite {Gidd} is just a gauge choice
which is needed to maintain the linearized approximation.
A detailed formalism was developed  in Refs. \cite {gt,Gidd}, and 
was reiterated in Refs. \cite {Csaki2} and \cite {Rubakov2}
for the particular case at hand,
so we won't be repeating it here. 
We just point out that 
the brane bending reveals a ghost field which is 
used in \cite {Csaki2} to cancel the unwanted
graviton polarizations. A most simple way to see this is 
as follows. Suppose that a brane with no matter is located at
$y=0$. After a matter source is introduced on the brane, its 
location is shifted to $y-\zeta(x)$, where $\zeta(x)$
is some response function determined by the source. 
Thus, matter couples to 4D fluctuations through the 
warp factor $A(y-\zeta(x))$.
Expanding $ A(y-\zeta(x))$ in powers of $\zeta$ one finds 
an additional coupling of $T^{\mu\nu}$ to  $\zeta\partial A$.
This is the coupling which effectively introduces  a ghost.
Indeed, let us introduces a source with energy-momentum tensor
\beq
T_{\mu\nu}~\equiv ~S_{\mu\nu}~\delta({\bar y})~.
\eeq 
When the brane is bent by the matter source, 
one may choose  new coordinates
${\bar x},~ {\bar y}$
using a gauge transformations (see for details 
\cite {gt,Gidd,Csaki2,Rubakov2}).
The induced metric on the brane  
takes the following form  in these coordinates:
\beq
{\bar h}_{\mu\nu}(x,0)  \propto \int d^4z \left (D_5(x,0; z, 0)(
S_{\mu\nu}(z)-{1\over 3} \eta_{\mu\nu} S^\alpha_\alpha (z)) - 
H \eta_{\mu\nu}D_4(x,z)
S^\alpha_\alpha(z) \right )~,
\label{ghost}
\eeq
where $D_5$ denotes the scalar part of the 5D graviton propagator,  
$D_4$ denotes that of a four-dimensional scalar and $H$ is some 
positive constant proportional to the square root of the 
bulk cosmological constant. 
The last term in this expressions is equivalent to a 
contribution of  a scalar ghost field. 
We would like to point out here that brane bending term  
does not cause any problem in 
the RS scenario. Moreover, it is needed for self-consistency 
of the RS model.  Recall that in the RS framework 
there is a massless graviton zero-mode with 2 physical degrees of freedom,
in addition there is an unphysical ``graviscalar'' which is gauge
dependent, plus there are massive spin-2 gravitons.
The ghost in the RS framework is explicitly canceled by an {\it
unphysical} ``graviscalar''. 
Therefore, one is left with the 2 physical polarizations of the 4D
massless graviton zero-mode. 
One might think of this as canceling a longitudinal photon
by $A_0$  ``ghost'' in the Gupta-Bleuler   quantization of QED.
This cancellation of unphysical states does not take place  if 
there is no localized zero-mode. Which is precisely what 
happens in \cite {Rubakov1}.
As we discussed above, states which mediate 4D gravity in this case 
are just massive 
spin-2 states. They have 5 degrees of freedom which are {\it all
physical}.
Two degrees of freedom corresponding to ``graviphotons'' 
decouple at low energies, as they couple derivatively to 
a conserved energy-momentum tensor. However, the
third physical scalar does not decouple.   The aim of the ghost 
present in \cite {Csaki2} is to compensate for this scalar. 
Thus, one is left with the theory which  has a manifest {\it ghost}
in the physical spectrum.
This ghost was used in  \cite {Rubakov2}  
to remove the problem of extra degrees of freedom from the 4D theory 
to large distances, where gravity, in this case,  
becomes scalar antigravity  due to the ghost.

However, the presence of a ghost indicates     
sickness of a theory at {\it any}  scales.
In particular, the ghost energy is
unbounded from below. Any theory which looks remotely like 
gravity is then completely
unstable when coupled to such a state.
This instability is most probably due to the fact that the background
in \cite {Rubakov1}
violates positive-energy conditions \cite {Witten}.
 
Since there are no known ways to remedy  theories with physical ghosts,
we are inclined to take a conservative point of view and require that
ghost contributions  should be canceled for a sensible model.
In this case the model can be made  free of ghosts, 
however, the gravity in 4D  becomes a 
tensor-scalar gravity and 
one goes back  to the problems  pointed out in Ref.
\cite {DGP}. 

One may wonder whether the ghost can
persist at large distances. This is a bit confusing, since it 
naively seems that
the model of Ref. \cite {Rubakov1}
should become flat-five-dimensional at large distances  in which case  
the second term in (\ref {ghost})  is clearly absent. 
However, the theory at hand is never a flat-five-dimensional one. 
Rather it is a flat-five-dimensional model  with a peculiar brane. 
This brane is a combination of
a positive and negative tension slices and from large distances looks
as a zero tension object.  However, regardless of the fact that
this is a zero-tension object, 
it brakes {\it maximally} translation invariance in
extra dimensions. As a result,
there is no  continuous limit in which the theory is 
flat-five-dimensional.

Another possibility is to make the ghost metastable, and decay at large
distances. Still, even in this case, 
it should admit a K\"allen-Lehman representation in terms
of massive  ghost states. Again, metastability does not 
suffice to ``exorcise'' the ghost.
Indeed, this may at most cure problems in single-ghost exchange 
amplitudes, but not in
amplitudes involving two or more ghosts. The instability associated 
with the fact that
the ghost energy is unbounded-below is just an example. 

The ghost formulation of the problems raised in \cite {DGP} 
makes us think that  they might be related to the lack of 
energy-positivity in this scenario \cite {Witten}. Probably, any
solution of the ghost problem must also cure the 
energy-positivity problem. 
In any event, 
this framework deserves further investigation and perhaps there are  
some unconventional solutions to the problems discussed above. 

\vspace{0.3cm}

{\bf Acknowledgments}

\vspace{0.2cm} 

We would like to thank C. Cs\'aki, J. Erlich, and T.J. Hollowood
for useful communications regarding the results of Ref. \cite {Csaki2}.
The work of G.D. is supported in part by a David and Lucile  
Packard Foundation Fellowship for Science and Engineering. G.G. is
supported by  NSF grant PHY-94-23002. 
M.P. is supported in part by NSF grant PHY-9722083.

\end{document}